\documentclass[12pt,epsbox]{article}
\usepackage{latexsym}
\usepackage{psfig,amssymb,epsf}
\usepackage[dvips]{graphicx}

\makeatletter
\@addtoreset{equation}{section}
\renewcommand{\theequation}{\thesection.\arabic{equation}}

\newcommand{\bC}{{\bf C}}
\newcommand{\bS}{{\bf S}}

\newcommand{\cN}{{\cal N}}

\newcommand{\cP}{{\cal P}}

\newcommand{\nn}{\nonumber \\}

\newcommand{\be}{\begin{equation}} \newcommand{\ee}{\end{equation}}
\newcommand{\bea}{\begin{eqnarray}} \newcommand{\eea}{\end{eqnarray}}

\newcommand{\gt}{\widetilde{g}}

\newcommand{\At}{\widetilde{A}_3}

\font\zfont = cmss10 
\newcommand{\ZZ}{\hbox{\zfont Z\kern-.4emZ}}

\ifx\TwoupWrites\UnDeFiNeD\else\target{\magstepminus1}{11.3in}{8.27in}
\source{\magstep0}{7.5in}{11.69in}\fi
\newfont{\fourteencp}{cmcsc10 scaled\magstep2}
\newfont{\titlefont}{cmbx10 scaled\magstep3}
\newfont{\authorfont}{cmcsc10 scaled\magstep1}
\newfont{\fourteenmib}{cmmib10 scaled\magstep2}
\skewchar\fourteenmib='177
\newfont{\elevenmib}{cmmib10 scaled\magstephalf}
\skewchar\elevenmib='177
\makeatletter
\newcommand\nonsequentialeqnum{
\@addtoreset{equation}{section}
\def\theequation{\arabic{section}.\arabic{equation}}}
\newif\ifp@bblock \p@bblocktrue
\newcommand\nopubblock{\p@bblockfalse}
\newcommand\topspace{\hrule height 0pt depth 0pt \vskip}
\newcommand\p@bblock{\begingroup \tabskip=\hsize minus \hsize
\baselineskip=1.5\ht\strutbox \topspace-2\baselineskip
\halign to\hsize{\strut ##\hfil\tabskip=0pt\crcr
\the\Pubnum\crcr\the\date\crcr}\endgroup}

\renewcommand\titlepage{\ifx\TwoupWrites\UnDeFiNeD\null
\vspace{-1.7cm}\fi
\vskip0.6cm
\ifp@bblock\p@bblock \else\hrule height 0pt \relax \fi}
\makeatother
\newtoks\date
\newtoks\Pubnum
\newtoks\pubnum
\newcommand{\frontpageskip}{\vspace{12pt plus .5fil minus 2pt}}
\renewcommand{\title}[1]{\frontpageskip
\begin{center}{\titlefont #1}\end{center}\par}
\renewcommand{\author}[1]{\frontpageskip\par\begin{center}
{\authorfont #1}\end{center}
\nobreak
}

\renewcommand{\thanks}[1]{\footnote{#1}}
\renewcommand{\abstract}{\par\frontpageskip\centerline{
\fourteencp Abstract}
\vspace{8pt plus 3pt minus 3pt}}

\begin{document}

\begin{titlepage}
\hfill
\vbox{
    \halign{#\hfil         \cr
           TAUP-2725-03 \cr
           hep-th/0305049  \cr
           } 
      }  
\vspace*{20mm}
\begin{center}
{\Large {\bf Probing Flavored Mesons of Confining Gauge Theories\\
by Supergravity  }} 
\vspace*{15mm}

{\sc Tadakatsu Sakai}
\footnote{e-mail: {\tt tsakai@post.tau.ac.il}}
and {\sc Jacob Sonnenschein}
\footnote{e-mail: {\tt cobi@post.tau.ac.il}}

\vspace*{1cm} 
{\it {Raymond and Beverly Sackler Faculty of Exact Sciences\\
School of Physics and Astronomy\\
Tel-Aviv University , Ramat-Aviv 69978, Israel}}\\ 

\vspace*{5mm}

\end{center}

\begin{abstract}
We incorporate massive flavored fundamental quarks in the supergravity
dual of ${\cal N}=1$ SYM by
introducing D7 brane probes to the Klebanov Strassler background. 
We find probe configurations that solve the D7 equations of motion. 
We compute the quadratic fluctuations of the D7 brane and extract
the spectrum of vector and pseudo scalar flavored mesons. 
The spectra found are discrete and
exhibit a mass gap of the order of the glueball mass.

\end{abstract}
\vskip 2cm

May 2003

\end{titlepage}

\setcounter{footnote}{0}

\newpage
\section{ Introduction} 
 
In the search for a supergravity model dual to a  ``realistic''
strong coupling gauge dynamics (for a review see \cite{Aharony}),
 one should be able to incorporate   
 quarks in the fundamental representation of   
 an $SU(N_c)$ gauge theory. 
Most of the known supergravity backgrounds  
   duals of confining four dimensional gauge theories    
  either do not incorporate quarks at all or admit quarks in the adjoint rather than the fundamental  representation 
\footnote{
Even  bifundamental quarks in models  
 of $SU(N)\times SU(N)$ turn into an  adjoint ( and a singlet)  once the symmetry group is broken to the diagonal $SU(N)$ group.}.   
 
Since the early days of strings it has been understood that fundamental quarks should correspond to open strings.  
In the modern era of closed  string theory this   obviously  calls for D branes.  
Certain basic objects of gauge theories like baryons \cite{Wittenbaryon}, instantons, monopoles,  domain walls\cite{PS} and others \cite{LS} 
were shown to correspond to wrapped D brane probes.    
It is thus natural to wonder,  whether one can consistently add D brane probes  to supergravity backgrounds  duals of  confining gauge theories, which will play the role of fundamental quarks.  
In case that $N_f$ the number of D brane probes is much smaller than
$N_c$, 
one can convincingly argue that the backreaction 
of the probe on  the bulk geometry is negligible.

It is well known that open strings between parallel $N_f$ D7 and $N_c$ D3 play the role of flavored quarks  
in the $SU(N_c)$ gauge theory on the D3 4d world volume gauge theory.
Karch and Katz \cite{KK;02} proposed to elevate this brane configuration into a  
supergravity background  
by introducing a D7 brane probe into the $AdS_5\times S^5$ background.

This idea was further explored in \cite{KKW}. In \cite{KMMW} 
the spectrum of an ${\cal N}=2$ $SU(N_c)$ SYM with fundamental
hypermultiplet was extracted from 
the supergravity background of $AdS_5\times S^5$ with D7 brane probe .
 It was found that for massive quarks the spectrum 
was discrete with a mass gap. They also analyzed the semiclassical rotating open strings attached to the D7 brane and discussed meson meson interaction.   
For a related work, see also \cite{d7}

The purpose of this paper is two folded: (i) to find a supergravity dual of
a confining gauge theory with flavors and  (ii) to  compute the mass
spectra of the pseudo and vector mesons made of the flavored quarks
from the SUGRA background.
The first goal  was  in fact proposed but not explicitly proven in
\cite{KK;02} and as for the second goal  
recall that the computation in \cite{KMMW} involves $\cN=2$ SQCD
which is not a confining gauge theory.
The main idea behind our work is to introduce D7 brane probes into the
Klebanov Strassler (KS) \cite{KS} background
(this idea was first given in \cite{KK;02}).
We will see that this configuration yields
an $\cN=1~SU(N_c)$ gauge theory with massive $N_f$ flavors.
The spectrum of the pseudo scalar and vector mesons can be
extracted from the spectrum of scalar and vector fluctuations of
the $N_f$ D7.
It turns out that it is advantageous not to use the formulation of the
deformed conifold of \cite{defconifold} but rather 
the coordinates introduced in \cite{cobi;hadron}. In the latter picture
there is a separation of the coordinates of the three manifold and
the $\bS^2$. 
It would be quite interesting to compare our results with a field
theory analysis. We will leave it as a future problem.

To understand the construction of the probe D7 branes recall that in 
the type IIA T dual picture the singular locus of  
the  conifold  takes the form of  a pair of perpendicular NS five 
branes \cite{ov,tconi}. Once the conifold is deformed there will be a 
``diamond'' structure at the intersection of the 
NS5 branes \cite{AKLM;99}. 
The fractional D3 branes of the KS model
are mapped by the T duality to D4 branes connecting the two NS branes. 
Now in this type IIA configuration one adds D6 branes on top of the NS5
branes. The strings between the D6 and the D4 branes play the role of
flavored fundamental quarks of the $SU(N)$ gauge theory\cite{EGK}.   
T dualizing back to IIB the D6 branes get mapped into D7 branes that intersect the singular locus. 
The D7 branes in our configuration 
span the world volume coordinates of the D3 branes, the radial
direction and wrap the  three cycle  
of the deformed the conifold. 
The transverse coordinates are the coordinates of the $\bS^2$ part
of the base.  
In terms of the transverse coordinates this means 
having one D7 brane at the north pole and another in the south pole of 
the $\bS^2$. These ``two'' D7 branes are in fact one D7 brane since they 
smoothly connect at the origin of the radial direction.    
By writing down the  8d effective action of the D7 branes which includes the DBI term and the WZ term,  
 we have shown that indeed these D7 brane 
configurations are solutions of the equations of motion.   
As a consistency check, 
the probe configuration satisfies the RR-flux cancellation condition. 
We will see that the probe configuration admits the isometry 
$SU(2)\times U(1)$ which is the subgroup of the isometry group of the
deformed conifold $SU(2)\times SU(2)$.

We then analyze the fluctuation modes around this classical D7 brane
solution.   
The worldvolume theory of the  D7 probe contains
$U(N_f)$ gauge fields as well as (pseudo) scalars in the adjoint
of the non-abelian flavor group.
Kaluza-Klein(KK) reduction of the fluctuations  
around the compact three manifold of the 8d  scalars yields 
5d (pseudo) scalars  
whereas the    
vector fields yield vectors as well as pseudo scalars. We then determine the quadratic fluctuations which are  
solutions of the equations of motion of the 5d effective action.  
Both type of fluctuations  have quantum numbers that are compatible with those of pseudo scalar and vector  
mesons of ${\cal N}=1$ SQCD, namely they are in the adjoint of $SU(N_f)$ and 
carry zero baryon number. To ensure that these modes are  
trivial under the $SU(2)\times U(1)$ global symmetry, 
we consider only a massless 5d vector 
fluctuation and the lowest scalar ones.    
  
A crucial point in the extraction of the mesonic spectra from the
corresponding fluctuations is that these 
are normalizable with an appropriate regularity condition satisfied
at the origin of the deformed conifold.
As a consequence, we show that the spectrum is discrete with a mass
gap for the vector mesons.  
We also argue that the pseudo scalar meson spectrum exhibits the same
behavior.
It is shown  that the system of mesons is stable (in the quadratic 
order approximation)   since 
 the kinetic term of the scalar  fluctuations is positive and since there are no tachyonic masses. 
The mass scale of the  mesons is that of the glueballs, namely,
$M_{\rm meson}\sim{\epsilon^{2/3}\over g_s N \alpha'}$.

The paper is organized as follows. 
In section 2, the KS background is described
 in the alternative formulation of \cite{cobi;hadron}. 
The action of the D7 probe including the DBI part and the WZ is analyzed
in section 3. We determine the solutions of the equations of motion that
correspond to two D7 
brane probes that merge smoothly into each other at the origin. 
Section 4 is devoted to the determination of the fluctuations around
the probe configurations and the extraction of the mesonic spectrum.  
In subsection 4.1 the quadratic fluctuations are analyzed and then 
in subsections 4.2 and 4.3 the spectrum of the vector mesons and the
scalar mesons is computed. We summarize and discuss some open questions
in section 5. Some useful formulae are summarized in the appendix.


\section{The KS model and the geometry of a deformed conifold}

The KS background is based on adding fractional D3 branes into the 
deform conifold.
The original KS metric   \cite{KS}
\begin{eqnarray}
ds_{10}^2=h^{-1/2}dx_{\mu}^2+h^{1/2}dx_6^2,
\label{ks}
\end{eqnarray}
made use of  the metric of a deformed conifold given in \cite{defconifold}.
It turns out that for our purposes it is more convenient to use the formulation of \cite{cobi;hadron}
since it admits a  separation between the three cycle and two cycle of
the deformed conifold. It is given by
\begin{eqnarray}
\epsilon^{-4/3}ds_6^2&\!=\!&
 {1\over 4}\,K(\tau)\,\cosh(\tau)\,\left(d\tau^2 + (\omega^a)^2\right) \nn
 &&+K(\tau)\,\sinh^2({\tau\over 2})\;
 \Bigg[\;
 (d\theta^2 + \sin^2\theta\,d\phi^2)
 -(\sin\phi\,\omega^1 + \cos\phi\,\omega^2)(d\theta) \nn
 &&\qquad - \,(\cos\theta\cos\phi\,\omega^1 - \cos\theta\sin\phi\,\omega^2
 - \sin\theta\,\omega^3)(\sin\theta\,d\phi) \nonumber
 \Bigg] \nonumber \\
&&+
 {1\over 4}\,K'(\tau)\,\sinh(\tau) \Big[\;d\tau^2\; +
 \; (\sin\theta\cos\phi\,\omega^1 +
 \sin\theta\sin\phi\,\omega^2
 +\cos\theta\,\omega^3)^2\Big]  .
\label{deformed}
\end{eqnarray}
Here
\begin{equation}
K(\tau)={(\sinh(2\tau)-2\tau)^{1/3}\over 2^{1/3}\sinh\tau},~~
h(\tau)=(g_sM\alpha^{\prime})^2\,2^{2/3}\,\epsilon^{-8/3}I(\tau),
\end{equation}
\begin{equation}
I(\tau)=\int_\tau^\infty dx\,{x\coth x-1\over \sinh^2\!x}
\left(\sinh(2x)-2x\right)^{1/3}.
\end{equation}
For instance it is easy to verify in this formulation that 
 for $\tau=0$, the 6d metric reduced to $\sum (\omega^a)^2$,
giving the $\bS^3$ while the $\bS^2$ shrinks to zero.

It is useful to rewrite the metric as follows
\begin{equation}
ds_{10}^2=R^2\left( m^2 I^{-1/2}dx_{\mu}^2+I^{1/2}d\tilde{x}_6^2\right),
\end{equation}
where
\begin{equation}
R^2=2^{1/3}g_sM\alpha^{\prime},~~
m^2={\epsilon^{4/3}\over 2^{2/3}(g_sM\alpha^{\prime})^2},
\end{equation}
and $\epsilon^{-4/3}ds_6^2=d\tilde{x}_6^2$.
$m$ is the mass scale of glueballs. For a detail of the glueball spectrum
in the KS background, see \cite{gball;ks}.

The NS B-field is
\begin{equation}
B_2={g_sM\alpha^{\prime}\over 2}\Big[
(f+k)\,\gt^3\wedge\gt^4+\sqrt{2}k\,\gt^3\wedge d\theta
+\sqrt{2}k\,\gt^4\wedge\sin\theta d\phi+2k\,\sin\theta d\theta\wedge d\phi
\Big]
.
\end{equation}
Here
\begin{eqnarray}
f(\tau)&\!=\!&{\tau\coth\tau-1\over 2\sinh\tau}\,(\cosh\tau-1) \ ,\nn
k(\tau)&\!=\!&{\tau\coth\tau-1\over 2\sinh\tau}\,(\cosh\tau+1) \ .
\end{eqnarray}
For the definition of the one-forms $\gt$, see the appendix.
The RR 2-form potential reads
\begin{equation}
C_2={M\alpha^{\prime}\over 2}\left[
{1\over 2}\cos\theta^{\prime}d\psi^{\prime}\wedge d\phi^{\prime}
+\sqrt{2}F(\tau)\left(\gt^3\wedge\sin\theta d\phi-\gt^4\wedge d\theta\right)
\right].
\end{equation}
Here
\begin{equation}
F(\tau)={\sinh\tau-\tau\over 2\sinh\tau} \ .
\end{equation}
We will later need the explicit expression for $C_6$ the RR six form. 
The corresponding RR 7-form field strength $F_7=dC_6$ is defined by
\begin{equation}
dF_7=d^{\ast}dC_2-F_5\wedge dB_2 \ ,
\end{equation}
where $F_5$ is the self-dual 5-form field strength given by
\begin{equation}
F_5={\cal F}_5+{}^{\ast}{\cal F}_5 \ ,
\end{equation}
with
\begin{eqnarray}
{\cal F}_5&\!=\!&B_2\wedge dC_2={gM^2\alpha^{\prime 2}\over 4}\,l(\tau)\,
g^1\wedge g^2\wedge g^3\wedge g^4\wedge g^5 \ , \nn
{}^{\ast}{\cal F}_5&\!=\!&{\alpha\,l(\tau)\over g_s\,K^2h^2\sinh^2\!\tau}\,
d^4x\wedge d\tau \ ,
\end{eqnarray}
with $\alpha=4(g_sM\alpha^{\prime})^2\epsilon^{-8/3}$ and
\begin{equation}
l(\tau)=f(1-F)+kF={\tau\coth\tau-1\over 4\sinh^2\!\tau}(\sinh 2\tau-2\tau) \ .
\end{equation}
>From the equation of motion of $C_2$ it follows that $d F_7=0$. 
We thus  obtain
\begin{eqnarray}
C_6&\!=\!&{M\alpha^{\prime}\over 4}(f^{\prime}+k^{\prime})h^{-1}
d^4x\wedge d\tau\left(-\sin\theta\cos\phi\,\omega^1
+\sin\theta\sin\phi\,\omega^2-2\cos\theta\,d\phi\right) \nn
&\!-\!&{M\alpha^{\prime}\over 4}\,(k-f)h^{-1}\,d^4x\left[
\sqrt{2}\left(\gt^3\,d\theta+\gt^4\sin\theta d\phi\right)
+2\sin\theta d\theta d\phi\right] \ .
\end{eqnarray}
where ${}^\prime=\partial/\partial\tau$.
Finally we notice that the RR-4-form gauge potential is defined by
\begin{equation}
F_5=dC_4+B_2\wedge dC_2=B_2\wedge dC_2+{}^{\ast}(B_2\wedge dC_2) \ ,
\end{equation}
from which we obtain
\begin{equation}
C_4={2^{4/3}(mR)^4\over g_s}\,u(\tau)\,d^4x \ .
\end{equation}
Here $u(\tau)$ is defined by
\begin{equation}
u^{\prime}(\tau)={l(\tau)\over K^2I^2\sinh^2\!\tau} \ .
\end{equation}

Let us now review briefly some geometrical aspects of the deformed
conifold.
The deformed conifold is defined by
\begin{equation}
w_1w_2-w_3w_4=-{\epsilon^2\over 2} \ .
\label{def;deformed}
\end{equation}
For the explicit form of $w_i$, see the appendix.
This can be regarded as a
$\bC^{\ast}$ fibration over a two-dimensional complex plane spanned
by $w_3,w_4$. The fibers get degenerate when $w_1w_2=0$,
giving a smooth singular locus in the base which is T-dual to
an NS5-brane \cite{ov}.
It turns out that the condition $w_1w_2=0$ can be solved as
(1) $\theta=0,~\theta^{\prime}=\pi$, (2) $\theta=\pi,~\theta^{\prime}=\pi$.
Each corresponds to a cylinder that intersect with one another at a
circle. To see this, note that for the two cases one finds
\begin{eqnarray}
&&w_1=w_2=0,\nn
&&w_3={i\epsilon\over\sqrt{2}}e^{{i\over 2}(\phi^{\prime}-\psi^{\prime})}
\left( \pm\sinh{\tau\over 2}-\cosh{\tau\over 2}\right), \nn
&&w_4={i\epsilon\over\sqrt{2}}e^{-{i\over 2}(\phi^{\prime}-\psi^{\prime})}
\left( \pm\sinh{\tau\over 2}+\cosh{\tau\over 2}\right).
\end{eqnarray}
These denote two cylinders spanned by $\tau,\phi^{\prime}-\psi^{\prime}$
that intersect with each other at a circle at $\tau=0$.
Recall that the circle is embedded in the $\bS^3$ and has a radius
proportional to $\epsilon$.
This circle corresponds to a diamond 
\cite{AKLM;99}.
In the limit $\epsilon\rightarrow 0$ where the deformed conifold reduces
to a conifold, the radius goes to zero
so that the smooth locus is split into two separate  cones that intersect
with each other at the tips of the cones.
The T-dual of the conifold \cite{tconi}
gives us the IIA brane configuration that
consists of two perpendicular NS5 branes, one at $w_1=0$ and the other
at $w_2=0$. 

\section{ The D7 brane probes in the KS background}

Now let us consider D7 probes in the KS background that yield massive
flavors, which were first discussed in \cite{KK;02}. 
For that purpose, it is useful to start from the T-dualized
picture of the conifold limit $\epsilon\rightarrow 0$.
The fractional D3-branes get mapped to D4-branes extending to
the cycle. 
In this setup, one can add D6-branes on top of either of the two
perpendicular NS5-branes
to get ${\cal N}=1$ SQCD with massless flavors \cite{BH},
because the D4-branes divide the D6-branes into two pieces, each of
which is responsible for chiral flavor groups.
Upon T-dualizing the IIA brane configuration back, each type of
D6-branes gets mapped to different D7-branes \cite{PRU}:
one is D7-branes which intersect with the singular locus $w_1=0$ and
the other D7-branes intersect with $w_2=0$.

\begin{figure}
\begin{center}
\includegraphics[scale=0.8]{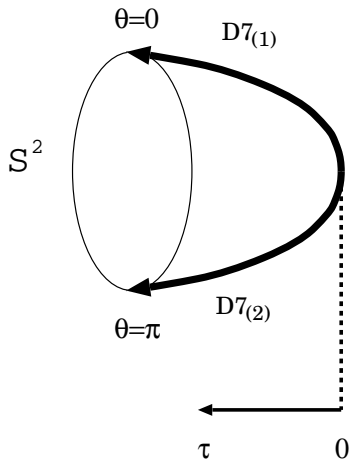}
\end{center}
\caption{\footnotesize{
Configuration of 7-branes on 2-sphere
}}
\label{s2}
\end{figure}

Now turn on $\epsilon\ne 0$ to have the deformed conifold.
As seen before, the singular loci become a single smooth cylinder.
Correspondingly, the two kinds of D7-branes intersecting the singular
loci become a smooth component of D7-branes that intersects
with the smooth locus.
More precisely, let us define
D7$_{(1)}$-branes staying at $\theta=0$, the north pole of the $\bS^2$ and 
D7$_{(2)}$-branes staying at $\theta=\pi$, the south pole of the $\bS^2$.
See figure \ref{s2}.

We show that
these two seven branes intersect with each other at the 
$\bS^3$ at $\tau=0$ and form a smooth D7-brane:
we first notice that the world volumes
of the D7$_{(1)}$ and D7$_{(2)}$ are charactorized by
\begin{eqnarray}
w_{1,2}^{(1)}&\!=\!&{\epsilon\over\sqrt{2}}\,\cos{\theta^{\prime}\over 2}
\,e^{\pm{i\over 2}(\phi^{\prime}+\psi^{\prime})}
\left(\sinh{\tau\over 2}\pm\cosh{\tau\over 2}\right) \ , \nn
w_{3,4}^{(1)}&\!=\!&{i\epsilon\over\sqrt{2}}\,\sin{\theta^{\prime}\over 2}
\,e^{\pm{i\over 2}(\phi^{\prime}-\psi^{\prime})}
\left(\sinh{\tau\over 2}\mp\cosh{\tau\over 2}\right) \ ,
\end{eqnarray}
and
\begin{eqnarray}
w_{1,2}^{(2)}&\!=\!&{\epsilon\over\sqrt{2}}\,\cos{\theta^{\prime}\over 2}
\,e^{\pm{i\over 2}(\phi^{\prime}+\psi^{\prime})}
\left(-\sinh{\tau\over 2}\pm\cosh{\tau\over 2}\right) \ , \nn
w_{3,4}^{(2)}&\!=\!&{i\epsilon\over\sqrt{2}}\,\sin{\theta^{\prime}\over 2}
\,e^{\pm{i\over 2}(\phi^{\prime}-\psi^{\prime})}
\left(-\sinh{\tau\over 2}\mp\cosh{\tau\over 2}\right) \ .
\end{eqnarray}
It is easy to see that
\begin{equation}
\omega_i^{(1)}(\tau=0)=\omega_i^{(2)}(\tau=0) \ , ~~~
\partial_{\tau}\omega_i^{(1)}(\tau=0)=
-\partial_{\tau}\omega_i^{(2)}(\tau=0) \ .
\end{equation}
These relations guarantee the smooth connection of the 7-branes at
$\tau=0$.

Recall that in the deformed conifold the chiral flavor symmetry is
broken to the diagonal subgroup $U(N_f)$, as we are left with only a
single type of D7-branes.
This implies that turning on $\epsilon$ amounts to turning on a vev
for a Higgs field on the 7-branes that transforms as the bifundamental
representation of the chiral flavor group.
This vev plays the role of a mass term in the 4d gauge theory on the
fractional D3 branes.


Let us now analyze
the D7-probe action in the KS background to verify that
the configuration solves the equation of motion.
The action consists of two parts
\begin{eqnarray}
S=S_{\rm DBI}+S_{\rm WZ} \ ,
\end{eqnarray}
where
\begin{eqnarray}
S_{\rm DBI}&\!=\!&-{\mu_7\over g_s}\int d^8\sigma\,\sqrt{
-\det\left(\phi^{\ast}g+\phi^{\ast}B_2+2\pi\alpha^{\prime}F\right)} \ ,\\
\label{DBI}
S_{\rm WZ}&\!=\!&-\mu_7\int e^{2\pi\alpha^{\prime}F+B}\wedge\sum_p C_{p+1} \nn
&\!=\!&-\mu_7\int\left(
{1\over 6}(2\pi\alpha^{\prime}F+B)^3C_{2}
+{1\over 2}(2\pi\alpha^{\prime}F+B)^2C_{4}
+(2\pi\alpha^{\prime}F+B)C_6 \right) \ .
\label{WZ}
\end{eqnarray}
Here $\phi^{\ast}g,\phi^{\ast}B_2$ are the pull-backs of $g,B_2$.
$F$ is the $U(1)$ field strength on the brane.
$\mu_7=1/(2\pi)^{7}\alpha^{\prime 4}$.
The two transverse coordinates of a probe D7 are taken to be
$\theta,\phi$ 
while
upon taking the static gauge the world volume coordinates are
given by $X^{\alpha}=(X^I,\theta^{\prime},\phi^{\prime},\psi^{\prime})$,
with $X^I=(x^{\mu},\tau)$.
What we should show is that the D7$_{(1)}$ given by
$\theta=0,~{\rm any}~\phi$, and the D7$_{(2)}$ given by
$\theta=\pi,~{\rm any}~\phi$ solve the equation of motion.
In order to check first the D7$_{(1)}$ brane,
we assume $\theta={\rm const},~\phi=\phi(X^{\alpha})$ and expand the
action around $\theta=0$.
It is verified that the coefficients of the linear term in $\theta$
depend linearly on some of the components of $F$.
By requiring them to vanish, we find that the $U(1)$ gauge
potential on the probe should take the form
\begin{equation}
A=A_3(\tau)\omega^3 \ .
\label{probe;gauge}
\end{equation}
It is then not difficult to show that the action becomes
\begin{eqnarray}
S_{\rm DBI}&\!=\!&-{\mu_7R^8\over g_s}
\int d^8X\sqrt{-\det(g_{(8)}+R^{-2}B_2^{(0)})} \nn
&&\quad
\times\Bigg[\left(1+{4\over I^{1/2}(K\sinh\tau)^{\prime}}\,
g^{\tau\tau}\partial_\tau\At\,\partial_\tau\At\right)
\left( 1-{8x\over I^{1/2}J}\At+{16\over IJ}\At^2\right)
\Bigg]^{1/2}\Big(1+{\cal O}(\theta^2)\Big) \ , \nn\nn
S_{\rm WZ}&\!=\!&-{\mu_7R^8m^4\over 2^{4}g_s}\int d^4x\,d\tau\,
\omega^1\omega^2\omega^3 \,\Big(1+{\cal O}(\theta^2)\Big)\nn
&&\times\Bigg[
2^{-2/3}(f+k)(f^{\prime}+k^{\prime}) I^{-1}
+\Big(2^{5/3}(f^{\prime}+k^{\prime}) I^{-1}+2^3\,\partial_{\tau}((f+k)u)\Big)
\,\At -2^{13/3}\,u^{\prime}\,\At^2
\Bigg] \ . \nn
\label{Sprobe}
\end{eqnarray}
Here $\At={2\pi\alpha^{\prime}\over R^2}A_3$ and
\begin{eqnarray}
g_{(8)\alpha\beta}dX^{\alpha}dX^{\beta}&\!=\!&g_{IJ}dX^IdX^J
+{I^{1/2}\over 4}\Big( K\cosh\tau\left( (\omega^1)^2+(\omega^2)^2\right)
+(K\sinh\tau)^{\prime} (\omega^3)^2\Big) \ , 
\label{8dmetric}
\\
R^{-2}B_{2}^{(0)}&\!=\!&-{f+k\over 2^{7/3}}\,\omega^1\wedge\omega^2 \ ,
\end{eqnarray}
with
\begin{equation}
g_{IJ}dX^IdX^J=m^2I^{-1/2}dx_{\mu}^2
+{I^{1/2}\over 4}\,(K\sinh\tau)^{\prime}\,d\tau^2 \ .
\label{5dmetric}
\end{equation}
Also
\begin{equation}
J=(K\cosh\tau)^2+x^2 \ ,~~~
x={f+k\over 2^{4/3}I^{1/2}} \ .
\end{equation}
Since there appear no linear terms in $\theta$, we find that 
$\theta=0$ solves the equation of motion of $\theta$.
However $A_3=0$ is not a solution because 
$A_3$ couples to a non-trivial current.
We can easily see that the current is conserved.
Recall that the contribution of the current from the DBI action is due
to the non-trivial NS B-fields.\footnote{
The backgrounds discussed in \cite{KK;02} have no non-trivial
B-fields and hence the gauge field configurations on the probe
branes are trivial.}
The equation of motion of $A_3$ is so complicated that 
no explicit form of the solution is found.
The solution must be given to evaluate the tension of
the probe 7-brane. 
We leave it as an open problem.
However, the D7$_{(1)}$ probe configuration itself is allowed as
the existence of the solution to the equation of motion
should be guaranteed.
It is interesting that the gauge configuration induces a D3-brane charge
on the probe:
\begin{equation}
\int F\wedge F=2 \int d\tau \,\omega^1\omega^2\omega^3 
\,F_{\tau 3}F_{12} \ .
\end{equation}

Also by expanding the action around 
$\theta=\pi$ with any $\phi$ and the same form of the $U(1)$ potential,
we obtain the same form of the action as above.
Hence the D7$_{(2)}$ probe also solves the equation of motion.

Recall that (\ref{8dmetric}) gives the metric of the 8d hypersurface
that denotes the embedding of the D7$_{(1)}$ and D7$_{(2)}$ probes
in KS.
We will refer to the compact 3d submanifold as $M_3$. 
$M_3$ is topologically
$\bS^3$ and admits the isometry group $SU(2)\times U(1)$, where
$SU(2)$ is the left transformation of $T$ that leaves $\omega^a$
invariant and $U(1)$ is the subgroup of the right $SU(2)$ under which
$\omega^{1,2}$ transform as a doublet while $\omega^3$ is invariant.
One finds that the isometry group leaves $B_2,C_2,C_4$ and $A$
invariant as well.

Here it is crucial to notice that the solution satisfies the consistency
condition of the RR-flux cancellation.
>From the 10d point of view the two D7 branes carry opposite RR charges
on $\bS^2$ so that the total charge vanishes. 
See figure \ref{s2}.

\begin{figure}
\begin{center}
\includegraphics[scale=1]{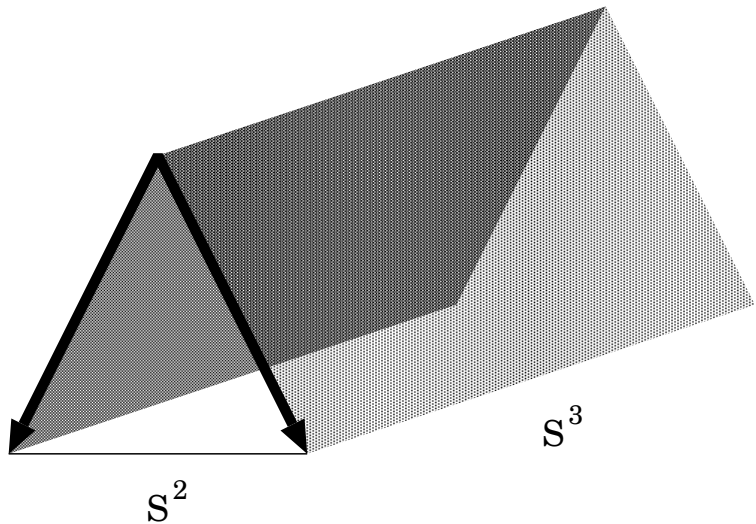}
\end{center}
\caption{\footnotesize{
D7-probe configuration in a deformed conifold. 
The vectors denote the singular locus
where the elliptic fibers get degenerate. The two shadowed surfaces that
intersect with each vector are the 7-brane probes.}}
\label{d7}
\end{figure}

To summarize, the probe D7-brane configuration looks like figure \ref{d7}.

\section{Spectrum of mesons}

In this section, we study the fluctuation modes around
the D7 probe configuration.
Since these belong to the adjoint representation
of $U(N_f)$, so do the dual states in the QCD.
We thus expect to find the information about 
mesons made out of the fundamental  quarks. Recall that no state in the adjoint of  $U(N_f)$
is charged under the baryon number $U_B(1)$,  where $U(N_f)=SU(N_f)\times U_B(1)$.
There exist two kinds of  mesons we can learn about  in the present
context: pseudo-scalar and vector mesons.
To see this, let us start with the D7 action defined on
the probe configuration. We have two kinds of dynamical fields on it,
 8d gauge potentials and
8d scalars that correspond to the fluctuation along
the transverse directions.
Upon KK reduction around $M_3$, we obtain an effective 5d theory that
consists of the infinite number of 5d vector and 5d scalar fields.
We see that the 5d vectors are dual to vector mesons in the dual QCD,
and the 5d scalars dual to pseudo-scalar mesons.
Here we consider only the lowest-lying KK modes,
namely, a massless 5d vector and two 5d scalar fields,
one coming from KK reduction of a 8d scalar by a scalar harmonics
and the other from that of a 8d vector by a vector harmonics.
The reason why we ignore the other massive modes is that
these modes are irrelevant to the physics of QCD:
since $M_3$ has the isometry $SU(2)\times U(1)$, 
the nontrivial harmonics carry 
non-zero spins and $U(1)$ charges. 
However there is no counterpart of these quantum numbers in the dual QCD.

In this section, we examine quadratic fluctuations of the D7 action
with $N_f=1$
around the probe configuration and derive the effective 5d theory
that involves only the lowest-lying KK modes.
As far as our concern is in quadratic fluctuations, we can concentrate
on the case $N_f=1$, since to this order any non-abelian interaction is
irrelevant and consequently a field in the adjoint representation of
$U(N_f)$ reduces to $N_f^2$ free fields.
Based on this, we perform a numerical computation of the spectrum
of vector and scalar mesons. In fact these are pseudo scalars rather than scalars as
follows from a straightforward parity analysis.
Recall that the 8d super Yang-Mills(SYM)
theory on a D7 comes from dimensional reduction of 10d SYM.
In ten dimensions, the parity transformation for the 10d vector field
is defined as
\begin{eqnarray}
\cP A(t,\vec{x}_9)\cP^{-1}=-A(t,-\vec{x}_9) \ .
\end{eqnarray}
Upon dimensional reduction to 8d, we find that the 8d scalar
and vector fields have parity odd.
Since the scalar and vector harmonics on $M_3$ have parity even,
the 5d scalars and vectors have parity odd.

\subsection{Quadratic perturbations around the probe configuration}

Here we consider the fluctuation around the D7$_{(1)}$-branes
in some detail.
The analysis of the D7$_{(2)}$-brane is then straightforward.
In fact, one ends up with the same results in the two cases.

As mentioned above, we assume
\begin{equation}
\theta=\theta(X^I),~~\phi=\phi(X^I) \ .
\label{assump}
\end{equation}
The 8d scalars $\theta$ and $\phi$ reduce to the 5d scalars
via a constant mode on $M_3$. 
The constant mode is the normalizable zero mode of the scalar
harmonics as $M_3$ is compact. 
The 8d vector $A$ yields a 5d massless vector and a 5d scalar
field.
To see which vector harmonics is to be taken to obtain the 5d scalar, 
we first recall that
the lowest-lying vector harmonics on
$\bS^3$ transform under the isometry $SU(2)\times SU(2)$ as 
$({\bf 3},{\bf 1})\oplus ({\bf 1},{\bf 3})$.
$M_3$ at hand is topologically $\bS^3$ and admits as the isometry
the subgroup $SU(2)\times U(1)$. The lowest-lying vector
harmonics then splits into some of the vector harmonics on $M_3$ 
that belong to an irreducible representation of $SU(2)\times U(1)$,
one of them being singlet. In fact, it is found that the singlet vector
harmonics is given by $\omega^3$. 
We thus find that the the 5d scalar corresponds to the
fluctuation of $A_3$ in (\ref{probe;gauge}) around the solution of
the system (\ref{Sprobe}).
However, since the solution is not known, one can not
obtain the explicit form of the action that governs the fluctuation.
As we will see below, up to quadratic order this
fluctuation mode decouples from the other 5d scalar and vector modes.
In this paper, we focus on only the two modes to compute the meson
spectrum.

It turns out that up to quadratic terms in the fluctuations, the DBI
action takes the form
\begin{eqnarray}
S_{\rm DBI}&\!=\!&-{\mu_7R^8m^4\over 2^4\,g_s}\int d^5X\,
\omega^1\omega^2\omega^3\,
(K\sinh\tau)^{\prime}J^{1/2} \nn
&&
\times \Bigg[1+{L\over 2}\,g^{IJ}\partial_I\theta\partial_J\theta
+{1\over 4}\left({2\pi\alpha^{\prime}\over R^2}\right)^2
g^{IJ}g^{KL}F_{IK}F_{JL}
\nn
&&~~~~ 
+\left({2\pi\alpha^{\prime}\over R^2}\right)^2\left(
{1\over 2}\,g^{IJ}u^{33}
\,\partial_IA_3\,\partial_JA_3 
+{8(K\cosh\tau)^2 \over IJ^2}\,A_3^2 \right)
-{2\pi\alpha^{\prime}\over R^2}\,{8x\over I^{1/2}J}A_3
\Bigg] \ .
\label{fluc}
\end{eqnarray}
Here
\begin{equation}
L={K\cosh\tau\over 4I^{1/2}J}\left[IK^2\sinh^2\!\tau
+{1\over 2^{8/3}}(\tau\coth\tau-1)^2\right] \ .
\end{equation}
$a,b=1,2,3$ are the indices of the orthonormal frame of $\bS^3$,
see the appendix for detail.
$u_{ab}$ is defined by the 3d part of the tensor 
$g_{(8)}+R^{-2}B_2^{(0)}$ and given by
\begin{eqnarray}
u_{ab}={I^{1/2}\over 4}\left(
\begin{array}{ccc}
K\cosh\tau & -x & 0 \\
+x & K\cosh\tau & 0 \\
0 & 0 & (K\sinh\tau)^{\prime}
\end{array}
\right) \ .
\end{eqnarray}

On the other hand, 
we find that the fluctuations from the WZ term up to quadratic order
are given by $S_{\rm WZ}$ in (\ref{Sprobe}).
This shows that $\theta$ and $A_I$ have no mixing terms.
As mentioned before, $A_3$ can not be diagonalized because
one has to expand this around a non-trivial configuration that solves
the action (\ref{Sprobe}).

It is interesting to notice that the kinetic term of $\theta$ is always
positive. This implies the stability of the probe configuration at hand.

\subsection{Vector mesons}

Let us first study the 5d vector part to compute the
vector meson spectrum.
Part of the analysis given here is parallel to \cite{dhr}
that aimed at investigating a localization of a bulk gauge field
at a brane world.

The action of the 5d gauge potential is
\begin{equation}
S_{\rm v}\sim m^4\int d^5X\,(K\sinh\tau)^{\prime}\,\sqrt{J}\,
g^{IJ}g^{KL}F_{IK}F_{JL}.
\end{equation}
Here we neglect the overall numerical factor because this is irrelevant
for the purpose of computing the mass spectrum of mesons.
The equation of motion reads
\begin{equation}
\partial_J\Big( (K\sinh\tau)^{\prime}\,\sqrt{J}\,g^{IJ}g^{KL}F_{IK}
\Big)=0.
\end{equation}
For the component $L=\tau$, this becomes
\begin{equation}
0=\eta^{\mu\nu}\partial_{\mu}(\partial_{\nu}A_{\tau}-\partial_{\tau}A_{\mu}).
\label{eom;tau}
\end{equation}
It is useful to work in the gauge $A_{\tau}=0$.
Then the above equation gives us the Gauss law constraint.
We next 
decompose the rest components of the gauge potential in terms of
the complete set of functions $\chi_n(\tau)$:
\begin{equation}
A_{\mu}(x,\tau)=\sum_n A_{\mu}^{(n)}(x)\chi_n(\tau).
\label{expansion}
\end{equation}
$\chi$ satisfy a second-order differential equation we will present
below. Note that we are interested in only normalizable solutions.
Substituting the decomposition into the Gauss law constraint, we obtain
\begin{equation}
0=\sum_n\eta^{\mu\nu}\partial_{\mu}A_{\nu}^{(n)}(x)\,
\partial_{\tau}\chi_n(\tau).
\end{equation}
As we will see later, the constant mode is not normalizable so that
we find
\begin{equation}
\eta^{\mu\nu}\partial_{\mu}A_{\nu}^{(n)}=0.
\end{equation}
This implies that $A_{\mu}^{(n)}$ are Proca fields.
Substituting (\ref{expansion}) into the action, we obtain
\begin{eqnarray}
S_{\rm v}&\!=\!&\int d^4x\,d\tau
\,(K\sinh\tau)^{\prime}\,\sqrt{J} \nn
&&\sum_{n,m}\Big[I
\,\eta^{\mu\nu}\eta^{\rho\sigma}
F_{\mu\rho}^{(n)}F_{\nu\sigma}^{(m)}\chi_n(\tau)\chi_m(\tau) 
+{8m^2\over (K\sinh\tau)^{\prime}}
\,\eta^{\mu\nu}A_{\mu}^{(n)}A_{\nu}^{(m)}
\,\partial_{\tau}\chi_n(\tau)\,\partial_{\tau}\chi_m(\tau)
\Big] \ . \nn
\end{eqnarray}
Now we define $\chi_n$ as the solution of the differential equation
\begin{equation}
-{1\over\sqrt{\gamma}}\,\partial_{\tau}\left(
{\sqrt{\gamma}\over (K\sinh\tau)^{\prime}\,I(\tau)}\,\partial_{\tau}\chi_n
\right)=\lambda_n\,\chi_n(\tau),
\label{deq;chi}
\end{equation}
with the normalization condition given by
\begin{equation}
\int_0^{\infty}d\tau\,\sqrt{\gamma}\,\chi_n(\tau)\chi_m(\tau)=\delta_{n,m} \ .
\label{normalization;chi}
\end{equation}
Here
\begin{equation}
\sqrt{\gamma}=(K\sinh\tau)^{\prime}\,I(\tau)\,\sqrt{J(\tau)} \ .
\end{equation}
It then follows that the action becomes
\begin{equation}
S_{\rm v}=\sum_n\int d^4x\Big[
\eta^{\mu\nu}\eta^{\rho\sigma}F_{\mu\rho}^{(n)}F_{\nu\sigma}^{(n)}
+8m^2\lambda_n\eta^{\mu\nu}A_{\mu}^{(n)}A_{\nu}^{(n)}\Big].
\end{equation}
Thus the Proca fields $A$ satisfy the on-shell condition with the mass
square given by
\begin{equation}
M_n^2=4\lambda_nm^2.
\end{equation}
We regard this as the mass square of vector mesons of QCD.

The differential equation (\ref{deq;chi}) allows two independent
solutions: one is normalizable and the other non-normalizable.
We are interested in the normalizable solution here.
It turns out that the normalizable solution should behave as
\begin{equation}
\chi(\tau)=e^{-{2\tau\over 3}}f(\tau),
\end{equation}
with the boundary behavior of $f$
\begin{equation}
f(\tau\rightarrow\infty)={\rm const}.
\end{equation}
As a check, we notice that the einbein $\sqrt{\gamma}$ behaves as
\begin{equation}
\sqrt{\gamma}\,(\tau\rightarrow\infty)\sim\tau^{3/2} \ .
\end{equation}
This guarantees the normalization condition (\ref{normalization;chi}).

It is easy to verify that $f$ obeys the differential equation
\begin{equation}
f^{\prime\prime}(\tau)+A(\tau) f^{\prime}(\tau)
+\big(B(\tau)+\lambda C(\tau)\big)f(\tau)=0,
\label{deq2;chi}
\end{equation}
with
\begin{eqnarray}
A&\!=\!&{1\over 2}\,\partial_{\tau}\log J-{4\over 3} \ ,\nn
B&\!=\!&-{1\over 3}\,\partial_{\tau}\log J+{4\over 9} \ ,\nn
C&\!=\!&(K\sinh\tau)^{\prime}\,I(\tau) \ .
\label{ABC}
\end{eqnarray}
We will solve this differential equation numerically following the
procedure discussed in \cite{glueball} to find the glueball spectrum
of QCD. 
We first find out the asymptotic behavior of the solution at $\tau \gg 1$
for a generic $\lambda$. Using this  data as an input, 
the solution can be found  numerically. By imposing a regularity
condition at $\tau=0$ to be discussed in a moment, only solutions
with appropriate values of $\lambda$  are allowed.
In particular, we will see that the spectrum of vector mesons is
discrete with a mass gap.

In order to obtain the asymptotic solution, we notice that for large $\tau$
\begin{eqnarray}
A(\tau)=\sum_{n=0,1,\cdots}a_n(\tau)e^{-2n\tau},~~
B(\tau)=\sum_{n=0,1,\cdots}b_n(\tau)e^{-2n\tau},~~
C(\tau)=e^{-{2\tau\over 3}}\sum_{n=0,1,\cdots}c_n(\tau)e^{-2n\tau},
\label{asym;abc}
\end{eqnarray}
where
\begin{eqnarray}
&&a_0=-{2\over 3},~~a_1={50\over 3},~~a_2={8\tau\over 3},~\cdots \nn
&&b_0=0,~~b_1=-{100\over 9},~~b_2=-{16\tau\over 9},~\cdots \nn
&&c_0=\tau,~~c_1={4\tau^2\over 3},~~c_2={32\tau^3\over 9},~\cdots
\end{eqnarray}
Expanding $f(\tau)$ as follows
\begin{equation}
f(\tau)=\sum_{n=0,1,\cdots}f_n(\tau)e^{-{2n\tau\over 3}},
\end{equation}
it follows from (\ref{deq2;chi}) that the coefficients $f_n$ obey the
recursion relation
\begin{eqnarray}
&&f_n^{\prime\prime}-{4n\over 3}f_n^{\prime}+{4n^2\over 9}f_n
+\sum_{m=0}^{\left[{n\over 3}\right]}\left[
a_m\,\left( f_{n-3m}^{\prime}-{2\over 3}\,(n-3m)\,f_{n-3m}\right)
+b_mf_{n-3m}\right] \nn
&&~~~~~~~~~~~~~~~~~~~~~~~~~~
+\lambda\sum_{m=0}^{\left[{n-1\over 3}\right]}c_mf_{n-3m-1}=0 \ .
\end{eqnarray}
By setting $f_0=1$, the solution is given by
\begin{equation}
f_0=1,~~f_1=-{9\lambda\tau\over 8},~~
f_2={27\lambda^2\tau^2\over 64},~~
f_3={25\over 12}-{81\lambda^3\tau^3\over 1024},~\cdots
\end{equation}

Now let us discuss what is the regulatory condition to be imposed
at $\tau=0$.
As seen before, the probe D7 brane consists of the two pieces,
D7$_{(1)}$ and D7$_{(2)}$. So far we have examined the 5d massless
vector potential only on one of the two 7-branes. 
The true solution is given by
interpolating smoothly between the two solutions each of which 
solves (\ref{deq;chi}).
We denote $\chi_n^{(1)}(\tau),~\chi_n^{(2)}(\tau)$ by the solutions
on D7$_{(1)}$ and D7$_{(2)}$, respectively.
Since both obey the same differential equation
with the same asymptotic behavior, the two solutions are related as
\begin{equation}
\chi_n^{(1)}(\tau)=\pm\chi_n^{(2)}(\tau).
\end{equation}
This shows that the regularity condition is given by
\begin{equation}
\partial_{\tau}\chi_n(\tau=0)=0~~{\rm or}~~\chi_n(\tau=0)=0.
\end{equation}
It follows from the regularity condition of vanishing of the derivative of $\chi_n$  that
the allowed values of the eigenvalue $\lambda$ are
\begin{equation}
\lambda=0.50,~1.64,~3.66,~\cdots
\end{equation}
which give us the mass spectrum of vector mesons
\begin{equation}
{M^2\over m^2}=2.00,~6.56,~14.6,~\cdots.
\end{equation}
and from  the regularity condition  of vanishing $\chi_n$ that
the values of  $\lambda$ are
\begin{equation}
\lambda=~0.86, ~2.50,~4.98,~\cdots
\end{equation}
which corresponds to  vector mesons masses
\begin{equation}
{M^2\over m^2}=~3.44,~10.0,~19.9,~\cdots.
\end{equation}

\subsection{Pseudo scalar mesons}

In this subsection, we compute the scalar meson spectrum.
As discussed in the previous subsection, there exist two scalar modes
in the 5d effective
action that correspond to the spectrum of scalar mesons that carries
no $SU(2)\times U(1)$ quantum numbers:
one is $\theta$ in (\ref{fluc}), and the other is $A_3$.

Let us first consider $\theta$.
As shown in (\ref{fluc}), the fluctuation on both of the
probe 7-branes is governed by
\begin{equation}
S_{\rm sc}\sim m^{2}\int d^4x d\tau (K\sinh\tau)^{\prime}\sqrt{J}\,
\,{L\over 2}g^{IJ}\partial_I\theta\,\partial_J\theta \ .
\end{equation}
We first decompose $\theta(X^I)$ 
in terms of the complete set of appropriate functions $\xi_n(\tau)$.
\begin{equation}
\theta(x^{\mu},\tau)=\sum_n\theta^{(n)}(x)\,\xi_n(\tau),
\end{equation}
By defining $\xi$ as the solution of
\begin{equation}
-{1\over\sqrt{\rho}}\,\partial_{\tau}\left(
{\sqrt{\rho}\over (K\sinh\tau)^{\prime}I}\,\partial_{\tau}\xi_n
\right)=\alpha_n\xi_n \ ,
\label{deq;xi}
\end{equation}
with the normalization condition given by
\begin{equation}
\int_0^{\infty}d\tau\sqrt{\rho}\,\xi_n(\tau)\,\xi_m(\tau)=\delta_{nm},
\end{equation}
where
\begin{equation}
\sqrt{\rho}=(K\sinh\tau)^{\prime}\sqrt{IJ}\,L,
\end{equation}
the action becomes
\begin{equation}
S_{\rm sc}={1\over 2}\int d^4x\sum_n\Big[
\eta^{\mu\nu}\partial_{\mu}\theta^{(n)}\partial_{\nu}\theta^{(n)}
+4\alpha_nm^2\theta^{(n)}\theta^{(n)}\Big].
\end{equation}
Thus we obtain the scalar mesons with the mass square
\begin{equation}
M^2_n=4\alpha_nm^2.
\end{equation}
In order to have the normalizable solutions, we see that that $\xi$
should behave as
\begin{equation}
\xi(\tau)=e^{-{4\tau\over 3}}g(\tau) \ ,
\end{equation}
with
\begin{equation}
g(\tau\rightarrow\infty)={\rm const}.
\end{equation}
As a check, we note that the einbein $\sqrt{\rho}$ behaves as
\begin{equation}
\sqrt{\rho}\,(\tau\rightarrow\infty)\sim \tau^{3/2}e^{2\tau/3} \ .
\end{equation}

We find that $g$ obeys the differential equation
\begin{equation}
g^{\prime\prime}(\tau)+D(\tau) g^{\prime}(\tau)
+\big(E(\tau)+\alpha C(\tau)\big)g(\tau)=0,
\label{deq2;xi}
\end{equation}
with
\begin{eqnarray}
D&\!=\!&\partial_{\tau}\log\left(
{\sqrt{\rho}\over (K\sinh\tau)^{\prime}\,I}\right)-{8\over 3} \ ,\nn
E&\!=\!&-{4\over 3}\,\left(D+{4\over 3}\right) \ .
\label{DE}
\end{eqnarray}
As before, we first solve the asymptotic behavior of $g$ for a 
generic $\alpha$. For that, we need the asymptotic behavior of the
coefficients $D,E$:
\begin{eqnarray}
D=\sum_{n=0,1,\cdots}e^{-2n\tau}d_n \ ,~~
E=\sum_{n=0,1,\cdots}e^{-2n\tau}e_n \ ,
\end{eqnarray}
where
\begin{eqnarray}
&&d_0=-{4\over 3} \ ,~~d_1={8\tau\over 3} \ ,~~
d_2={32\tau^2\over 3} \ , \cdots \nn
&&e_0=0 \, ~~e_1=-{32\tau\over 9} \ ,~~e_2=-{128\tau^2\over 9} \ ,
\cdots
\end{eqnarray}
Expanding $g(\tau)$ as
\begin{equation}
g(\tau)=\sum_{n=0,1,\cdots}g_n(\tau)e^{-{2n\tau\over 3}},
\end{equation}
it follows from (\ref{deq2;xi}) that the coefficients $g_n$ obey the
recursion relation
\begin{eqnarray}
&&g_n^{\prime\prime}-{4n\over 3}g_n^{\prime}+{4n^2\over 9}g_n
+\sum_{m=0}^{\left[{n\over 3}\right]}\left[
d_m\,\left( g_{n-3m}^{\prime}-{2\over 3}\,(n-3m)\,g_{n-3m}\right)
+e_mf_{n-3m}\right] \nn
&&~~~~~~~~~~~~~~~~~~~~~~~~~~
+\alpha\sum_{m=0}^{\left[{n-1\over 3}\right]}c_mg_{n-3m-1}=0 \ .
\end{eqnarray}
By setting $g_0=1$, the solution is given by
\begin{equation}
g_0=1,~~g_1=-{3\alpha\tau\over 4},~~
g_2={27\alpha^2\tau^2\over 128},~~
g_3={8\tau\over 15}-{81\alpha^3\tau^3\over 2560} \ ,~\cdots
\end{equation}
Using this, we solve (\ref{deq2;xi}) numerically.
As before, we have to impose the regularity condition at $\tau=0$:
\begin{equation}
\xi(\tau=0)=0 \ ,~~{\rm or}~~\partial_{\tau}\xi(\tau=0)=0 \ .
\end{equation}
We obtain from this 
\begin{equation}
\alpha=1.77,~3.91,~6.90,~\cdots
\end{equation}
which give us the mass spectrum of scalar mesons
\begin{equation}
{M^2\over m^2}=7.08,~15.6,~27.6,~\cdots.
\end{equation}

Unlike the vector  mesons, it turns out that for the pseudo scalar ones
the same values of $\alpha$ are found for both types of  the regularity
condition. This may be related to the fact that
around $\tau=0$ due to the shrinking of the 
$\bS^2$   the fluctuations and hence $\xi$ vanish  anyhow.

The computation of the meson spectrum associated with $A_3$
is left as an open question.

\section{Discussion}

In this paper, we have discussed adding D7-brane probes to the KS
background for the purpose of getting a SUGRA description of
${\cal N}=1$ SQCD with flavors.
The point here is that for $N_f\ll N_c$ the backreaction of the D7-branes
to the KS can be suppressed.
To find the probe configuration, the geometrical data of the deformed
conifold and its T-dual brane picture in IIA were useful.
Based on this observation, we discussed the configuration and verified
that it solves the equation of motion of the probe action.
As a consistency, we argued that the probe configuration satisfies
the RR-flux cancellation condition.
Using this result, we next computed the scalar and vector meson spectrum
by examining the normalizable fluctuations around the probe with an
appropriate regularity condition at $\tau=0$.
We discussed that the mass spectrum is
charactorized by the single mass scale $m$, being equal to
the glueball mass, and has a mass gap.
An important open problem here is to compute the scalar meson spectrum
that comes from $A_3$.

There still remain some issues to be explored.
One is to check whether  the probe configuration preserves ${\cal N}=1$
supersymmetry by examining $\kappa$-symmetry on the probe
brane, although that is plausible from the T-dualized IIA brane
picture.
It would be interesting to generalize the probe configuration
such that one is allowed to have one parameter family of the
solutions. For instance, D7 and D5 probe configurations
in $AdS_5\times S^5$ were found \cite{KK;02}
that depend on one parameter, being
identified with the mass of flavors.
It is also nice to compute Wilson lines via SUGRA to see how
the screening of a pair of heavy quarks in terms of dynamical quarks
occurs,
as was done for $AdS_5\times S^5$ with D7-probes \cite{KKW}.
Baryons could be realized as D3 brane probe which is wrapping the $M_3$
and is connected  with $N$ strings to the D7 brane probe.

In this paper, we have been working in the probe approximation, which
is justified for $N_f\ll N_c$.
In order to have a SUGRA configuration for any $N_f$, we need to find
a fully localized D7-brane configuration in the KS.
To achieve this sounds rather difficult, however. 
In fact, only a few example of fully localized brane solution are known
\cite{CH}.
Instead of working in the KS, it is useful to consider the Penrose limit
of the KS \cite{cobi;hadron}, where the original KS background gets much
simplified and the string spectrum on it is simple enough to work out
in light-cone gauge. 
One may expect to obtain a fully-localized D7-brane solution in the
resultant background.
It would be interesting to study the string
theory on that background.

As we have seen, the D7 probes yield massive flavors whose mass is
of order of the glueball mass.
Our model is not similar to a realistic QCD, unfortunately.
In fact, we are not allowed to take naively the limit 
$\epsilon\rightarrow 0$ to obtain massless flavors, because
this limit gives rise to a singularity in the KS and therefore
the SUGRA approximation is not valid any more.
It is quite interesting to find a probe configuration in an well-defined
SUGRA background that provides us with a more realistic model with
light quarks with chiral flavor symmetry.

\section*{Acknowledgements}
We would like to thank Yaron Oz and Stanislav Kuperstein for discussions. 
We would like to especially thank 
Ofer Aharony for useful conversations about the project and
for reading the manuscript.
This work was supported in part by the Israel Science Foundation
and the German-Israeli Foundation for Scientific Research and 
Development.

\appendix

\section{Formulae}

Here we summarize some useful relations.

Consider
\begin{equation}
T=e^{{i\over 2}\phi^{\prime}\sigma_3}\,
  e^{{i\over 2}\theta^{\prime}\sigma_1}\,
  e^{{i\over 2}\psi^{\prime}\sigma_3},~~
S=e^{{i\over 2}\phi\,\sigma_3}\,e^{-{i\over 2}\theta\,\sigma_1}.
\end{equation}
$T$ defines an $\bS^3$ as a Hopf fibration over $\bS^2$ spanned by
$\theta^{\prime},\phi^{\prime}$ and $S$ defines an $\bS^2$.
$\omega^a$ are defined as the left-invariant one-forms of the $\bS^3$:
\begin{equation}
T^{\dagger}dT={i\over 2}\,\omega^a\sigma_a.
\end{equation}
Explicitly
\begin{eqnarray}
w^1&\!=\!&\cos\psi^{\prime}d\theta^{\prime}
+\sin\theta^{\prime}\sin\psi^{\prime}d\phi^{\prime},\nn
w^2&\!=\!&\sin\psi^{\prime}d\theta^{\prime}
-\sin\theta^{\prime}\cos\psi^{\prime}d\phi^{\prime},\nn
w^3&\!=\!&d\psi^{\prime}+\cos\theta^{\prime}\sin\psi^{\prime}d\phi^{\prime}.
\end{eqnarray}
The left-invariant one-forms obey
\begin{equation}
d\omega^a={1\over 2}\epsilon^{abc}\omega^b\omega^c \ .
\end{equation}
The volume form is
\begin{equation}
w^1\wedge w^2\wedge w^3=\sin\theta^{\prime}\,
d\phi^{\prime}\wedge d\theta^{\prime}\wedge d\psi^{\prime} \ .
\end{equation}
The dual vectors are
\begin{eqnarray}
\partial_1&\!=\!&\cos\psi^{\prime}\,{\partial\over\partial\theta^{\prime}}
+{\sin\psi^{\prime}\over\sin\theta^{\prime}}
\,{\partial\over\partial\phi^{\prime}}
-\cot\theta^{\prime}\sin\psi^{\prime}
\,{\partial\over\partial\psi^{\prime}} \ , \nn
\partial_2&\!=\!&\sin\psi^{\prime}\,{\partial\over\partial\theta^{\prime}}
-{\cos\psi^{\prime}\over\sin\theta^{\prime}}
\,{\partial\over\partial\phi^{\prime}}
+\cot\theta^{\prime}\cos\psi^{\prime}
\,{\partial\over\partial\psi^{\prime}} \ , \nn
\partial_3&\!=\!&{\partial\over\partial\psi^{\prime}} \ .
\end{eqnarray}


In order to find the relation between the basis used in \cite{KS} and
that in the present paper, one first carries out a change of
basis for the 1-forms $\omega^i$ \cite{cobi;hadron},
\begin{equation}
  \sqrt{2}\,\tilde{g}^3\,\sigma_1-\sqrt{2}\,\tilde{g}^4\,\sigma_2+\tilde{g}^5\,\sigma_3 \;=\;
  S^\dag\,\omega^a\,\sigma_a\,S \ ,
\end{equation}
which gives
\begin{eqnarray}
  \gt^5 &=& \sin\theta\cos\phi\,\omega^1 - \sin\theta\sin\phi\,\omega^2
 + \cos\theta\,\omega^3, \nonumber\\
- \gt^4 &=& {1\over\sqrt{2}}\,(\sin\phi\,\omega^1 + \cos\phi\,\omega^2), \\
  \gt^3 &=& {1\over\sqrt{2}}\,(\cos\theta\cos\phi\,\omega^1 -
  \cos\theta\sin\phi\,\omega^2 - \sin\theta\,\omega^3). \nonumber
\end{eqnarray}
After a little algebraic work one finds
\begin{eqnarray}
  e^{-{i\over 2}\psi}\,(g^3 + i\,g^4) &=&  \gt^3 + i\,\gt^4   \nonumber \\
  g^5 &=& \gt^5  \\
  e^{-{i\over 2}\psi}\,(g^1 + i\,g^2) &=&  (\gt^3 -
  \sqrt{2}\sin\theta\,d\phi) + i\,(\gt^4 +
  \sqrt{2}\,d\theta) \nonumber
\end{eqnarray}

We next denote $w_i$ that defines the deformed conifold 
$w_1w_2-w_3w_4=-{\epsilon^2\over 2}$ in terms of $S$ and $T$.
Define
\begin{equation}
L=TS,~~R=\sigma_3\,S\,\sigma_3 \ ,
\end{equation}
from which, we define
\begin{equation}
W=L\,W_{\epsilon}R^{\dagger}=
\left(
\begin{array}{cc}
w_1 & w_3 \\
w_4 & w_2
\end{array}
\right),~~~
W_{\epsilon}={\epsilon\over\sqrt{2}}\,e^{{\tau\over 2}\sigma_3}\sigma_3 \ .
\end{equation}
This gives the deformed conifold:
\begin{equation}
\det W=-{\epsilon^2\over 2}.
\end{equation}
One can find that
\begin{eqnarray}
w_{1,2}&\!=\!&{\epsilon\over\sqrt{2}}\left[
\sinh{\tau\over 2}\cos\theta\cos{\theta^{\prime}\over 2}
e^{\pm{i\over 2}(\phi^{\prime}+\psi^{\prime})}
\mp i \sinh{\tau\over 2}\sin\theta\sin{\theta^{\prime}\over 2}
e^{\pm{i\over 2}(2\phi+\phi^{\prime}-\psi^{\prime})}
\pm\cosh{\tau\over 2}\cos{\theta^{\prime}\over 2}
e^{\pm{i\over 2}(\phi^{\prime}+\psi^{\prime})}
\right], \nn
w_{3,4}&\!=\!&{i\epsilon\over\sqrt{2}}\left[
\sinh{\tau\over 2}\cos\theta\sin{\theta^{\prime}\over 2}
e^{\pm{i\over 2}(\phi^{\prime}-\psi^{\prime})}
-\sinh{\tau\over 2}\sin\theta\cos{\theta^{\prime}\over 2}
e^{\pm{i\over 2}(2\phi+\phi^{\prime}+\psi^{\prime})}
\mp\cosh{\tau\over 2}\sin{\theta^{\prime}\over 2}
e^{\pm{i\over 2}(\phi^{\prime}-\psi^{\prime})}
\right]. \nn
\end{eqnarray}

\newpage


\end{document}